\documentclass[12pt,a4paper]{article}
\usepackage{amsmath, amssymb, amsfonts, mathrsfs, graphicx, fancyhdr,setspace}
\usepackage{hyperref}
\usepackage[square,sort,comma,numbers]{natbib}
\usepackage[pdftex]{color}
\def\be{\begin{equation}}
\def\ee{\end{equation}}
\def\ba{\begin{align}}
\def\ea{\end{align}}

\def\Dnote#1{{\bf [DC: #1]}}

\begin{document}

\renewcommand{\theequation}{\thesection.\arabic{equation}} 
\numberwithin{equation}{section} 
\onehalfspacing

\begin{titlepage}
\begin{center}
\rightline{\small DESY-15-162}
\vskip 1cm

{\Large \bf Classification of Shift-Symmetric No-Scale Supergravities}
\vskip 1.2cm

{\bf  David Ciupke$^{a}$ and Lucila Z\'arate$^{b}$}

\vskip 0.8cm

$^{a}${\em Deutsches Elektronen-Synchrotron DESY, Theory Group, D-22603 Hamburg, Germany}
\vskip 0.3cm

$^{b}${\em Fachbereich Physik der Universit\"at Hamburg, Luruper Chaussee 149, 22761 Hamburg, Germany}
\vskip 0.3cm

{\tt david.ciupke@desy.de, lucila.zarate@desy.de}

\end{center}

\vskip 1cm

\begin{center} {\bf Abstract } 
\end{center}

\noindent

Models of 4D $\mathcal{N}=1$ supergravity coupled to chiral multiplets with vanishing or positive scalar potential have been denoted as no-scale. Of particular interest in the context of string theory are models which additionally possess a shift-symmetry. In this case there exists a dual description of chiral models in terms of real linear multiplets. We classify all ungauged shift-symmetric no-scale supergravities in both formulations and verify that they match upon dualization. Additionally, we comment on the realizations within effective supergravities descending from string compactifications.

\vfill
\bigskip

\noindent September, 2015

\end{titlepage}

\newpage
\tableofcontents
\newpage

\section{Introduction}
No-scale models denote classes of matter-coupled supergravities in four dimensions 
with positive-/negative-semi-definite or vanishing scalar potentials \cite{Cremmer:1983bf, Barbieri:1985wq}. Originally these theories were conceived and studied for phenomenological purposes \cite{Ellis1984429}. Later on it was realized that no-scale supergravities arise as the low energy effective actions of certain superstring compactifications \cite{Witten:1985xb}. In particular, in compactifications of type II string theory the resulting 4D supergravity also enjoys a perturbative Peccei-Quinn shift-symmetry. Motivated by these examples, it is the purpose of this paper to classify all ungauged shift-symmetric no-scale supergravities.\footnote{In the following we will only regard ungauged models. A recent example of a gauged no-scale supergravity with shift-symmetry can be found in \cite{Dall'Agata:2013ksa}.}

For four-dimensional $\mathcal{N}=1$ supergravity coupled to a collection of chiral multiplets $T_i$ described via a shift-symmetric K\"ahler potential $K$ and superpotential $W$, the no-scale condition is defined as 
\begin{equation}\label{no_scale_intro}
  K^{T_i \bar{T}_{\bar{\jmath}}} K_{T_i} K_{\bar{T}_{\bar{\jmath}}} = p \ , \qquad \text{with} \quad p \in \mathbb{R} \ ,
\end{equation}
where $K_{T_i} = \partial K/ \partial T_i$ and $K^{T_i \bar{T}_{\bar{\jmath}}}$ denotes the inverse K\"ahler metric. We read this condition as a differential equation for $K$ and, hence,
the classification of all shift-symmetric no-scale supergravities amounts to solving this differential equation in full generality. 
As a first step towards this goal, we show that eq.~\eqref{no_scale_intro} is equivalent to the homogeneous Monge-Amp\`{e}re equation.\footnote{An alternative proof of this equivalence was also given in \cite{Barbieri:1985wq}.}
More precisely, due to the shift-symmetry one obtains the real version of the homogeneous Monge-Amp\`{e}re equation, which we then solve with standard methods of partial differential equations. The resulting solution is of a semi-explicit form and generalizes the solution for the special three-field case found in \cite{Chaundy}.\footnote{Note that, an alternative, fully implicit solution to the real homogeneous Monge-Amp\`{e}re equation was already given in \cite{Fairlie:1994in}.}

Furthermore, supergravity coupled to chiral multiplets enjoying the aforementioned Peccei-Quinn shift-symmetry has a dual description in terms of real linear multiplets \cite{Derendinger:1994gx, Binetruy:2000zx}. This dual description has been employed in the context of string compactifications where it often provides a more elegant framework to describe the dynamics of the shift-symmetric scalars, as for instance in \cite{Grimm:2004uq}. In a theory of linear multiplets the no-scale condition takes a simple form and in this paper we determine its general solution, which in contrast to the chiral case is explicit. Moreover, as a consistency check we demonstrate that the individual solutions to the no-scale condition in the formulation with chiral and linear multiplets match upon dualization.

In the second part of this paper we use the general, semi-explicit solution in the chiral formulation to construct explicit examples. In particular we recover the special class of models where $K$ is given as a logarithm of a homogeneous function. The no-scale property of such K\"ahler potentials was already pointed out in \cite{Ferrara:1994kg}. 
However, our general solution shows that K\"ahler potentials with such a homogeneity property form merely a special subclass among all possible explicit solutions. 
Thus, it is interesting to determine explicit, non-homogeneous models. For illustrative purposes we construct such solutions both of a more general type and for the special two-field case.

K\"ahler potentials with homogeneity property are particularly interesting in the context of stringy effective actions. We present an overview of such K\"ahler potentials descending from various compactifications of string theories. The homogeneity is an exact property of IIA/B orientifold flux-compactifications at tree level and persists even when including string tree-level perturbative $\alpha'$-corrections. It can also be found in compactifications of the heterotic string, but is confined there to the dilaton sector and large volume limit respectively. The homogeneity implies the existence of an additional Killing vector of the K\"ahler manifold related to dilatations. 
The associated scaling behaviour of the effective Lagrangian is an exact property of string theory at tree level as pointed out in \cite{Witten:1985xb}. This observation has two interesting consequences, the first being that a scaling symmetry is not a necessary feature of no-scale models. The second is that if stringy no-scale supergravities after inclusion of $g_s$-corrections exist, then they will most likely fall into the class of non-homogeneous functions, whose existence we demonstrated.
%

This paper is organized as follows. In sections \ref{chiral_multi} and \ref{lin_multi} we introduce the relevant notation of 4D $\mathcal{N}=1$ supergravity coupled to chiral and real linear superfields respectively. 
In section \ref{MAsol} we solve the homogeneous real Monge-Amp\`{e}re equation and, thus, classify shift-symmetric no-scale supergravities for chiral superfields. We derive the corresponding no-scale supergravities with linear superfields in section \ref{classification_linear}. In section \ref{explicitsolutions} we provide explicit examples of the solutions described above and make some remarks about the respective geometries. Finally in section \ref{noscale_string} we study the structure of no-scale models descending from string theory in detail. A proof of the equivalence between the Monge-Amp\`{e}re equation and the no-scale condition can be found in appendix \ref{MAeq}. In addition, in appendix \ref{duality_matching} we give the duality transformation between chiral and linear theories and explicitly show that the corresponding solutions match via dualization.
%
\section{Matter-Coupled $\mathcal{N}=1$ Supergravity in Four Dimensions}\label{mattercoupledsugra}
\subsection{Chiral Multiplets}\label{chiral_multi}
We begin by reviewing $\mathcal{N}=1$ supergravity coupled to a collection of chiral superfields $T_i$ , $i=1,\dots,n$ in four dimensions.\footnote{All the results of this paper directly extrapolate to theories with $\mathcal{N}=2$ supersymmetry in three dimensions.} In the following we partially adopt the notation and conventions of \cite{Binetruy:2000zx, Grimm:2005fa}. (Anti-) chiral superfields are defined by the condition
\begin{equation}
\bar{\mathcal{D}}^{\dot{\alpha}} T_i = \mathcal{D}_\alpha \bar{T}_{\bar{\jmath}} = 0 \ ,
\end{equation}
where $\bar{\mathcal{D}}^{\dot{\alpha}}$ and $\mathcal{D}_\alpha$ denote the covariant spinorial derivatives. The on-shell bosonic degrees of freedom of the chiral multiplets are complex scalars $T_i$, which parametrize a K\"ahler manifold $\mathcal{M}$ with K\"ahler potential $K(T_i,\bar{T}_{\bar{\jmath}})$. The coupling of the chiral superfields to supergravity can be conveniently performed in curved superspace via the Lagrangian 
\begin{equation}
 \mathcal{L} = -3 \int E + \frac{1}{2} \int \frac{E}{R} \mathrm{e}^{K/2} W(T_i)+ \frac{1}{2} \int \frac{E}{\bar{R}} \mathrm{e}^{K/2} \bar{W}(\bar{T}_{\bar{\jmath}}) \ ,
\end{equation}
where $W(T_i)$ is the superpotential, $R$ the curvature superfield and $E$ denotes the superdeterminant of the super-vielbein and implicitly depends on $K$.\footnote{The integration over the Grassmann variables is implicit in this notation.} The corresponding on-shell Lagrangian for the bosonic components reads
\begin{equation}\label{L:chiral}
 e^{-1} \mathcal{L} = - \tfrac{1}{2} \mathcal{R} - K_{T_i \bar{T}_{\bar{\jmath}}} \partial_{\mu} T_i \partial^{\mu} \bar{T}_{\bar{\jmath}} - V(T_i,\bar{T}_{\bar{\jmath}}) \ ,
\end{equation}
where $\mathcal{R}$ denotes the space-time scalar curvature, $e$ the determinant of the vielbein and $K_{T_i \bar{T}_{\bar{\jmath}}} = \frac{\partial}{\partial T_i} \frac{\partial}{\partial \bar{T}_{\bar{\jmath}}} K$ the K\"ahler metric. Furthermore the scalar potential is given by
\begin{equation}\label{chiralpot}
 V(T_i,\bar{T}_{\bar{\jmath}}) = \mathrm{e}^K ( K^{T_i \bar{T}_{\bar{\jmath}}} D_{T_i} W D_{\bar{T}_{\bar{\jmath}}} \bar{W} - 3 \lvert W \lvert^2 ) \ ,
\end{equation}
where $K^{T_i \bar{T}_{\bar{\jmath}}}$ describes the inverse K\"ahler metric and $ D_{T_i} W = \frac{\partial}{\partial T_i} W + W \frac{\partial}{\partial T_i} K$ the K\"ahler-covariant derivative. Defining the function
\begin{equation}\label{def_G}
 G(T_i,\bar{T}_{\bar{\jmath}}) = K(T_i,\bar{T}_{\bar{\jmath}}) + \mathrm{ln}(\lvert W(T_i) \lvert^2) \ ,
\end{equation}
we can recast the scalar potential into the form
\begin{equation}\label{PotInTermsOfG}
 V(T_i,\bar{T}_{\bar{\jmath}}) = \mathrm{e}^G ( G_{T_i} G^{T_i \bar{T}_{\bar{\jmath}}} G_{ \bar{T}_{\bar{\jmath}}} - 3)  \ ,
\end{equation}
where $G_{T_i} = \frac{\partial}{\partial T_i} G$ and $G^{T_i \bar{T}_{\bar{\jmath}}}$ denotes the inverse of $G_{T_i \bar{T}_{\bar{\jmath}}}= \frac{\partial}{\partial T_i} \frac{\partial}{\partial \bar{T}_{\bar{\jmath}}} G$.

In this paper we are interested in the special class of theories which are of the no-scale type. These are defined via the property 
\begin{equation}\label{eq:chiral_no_scale}
 G_{T_i} G^{T_i \bar{T}_{\bar{\jmath}}} G_{ \bar{T}_{\bar{\jmath}}} = p \ , \qquad p \in \mathbb{R} \ .
\end{equation}
From eq.~\eqref{PotInTermsOfG} it follows that these theories satisfy
\begin{equation}\label{noscalepot_cases}
 V > 0 \quad   \mathrm{for} \quad p > 3 \ , \qquad V = 0 \quad  \mathrm{for} \quad p = 3 \qquad \mathrm{and} \qquad V < 0 \quad \mathrm{for} \quad p < 3 \ .
\end{equation}
Note that all such models are related via a rescaling of the function $G$. 
Besides eq.~\eqref{eq:chiral_no_scale} one sometimes finds a second notion of no-scale models. This corresponds to a no-scale type condition on the K\"ahler potential, i.e.
\begin{equation}\label{weaknoscale}
 K_{T_i} K^{T_i \bar{T}_{\bar{\jmath}}} K_{ \bar{T}_{\bar{\jmath}}} = p \ , \qquad p \in \mathbb{R} \ .
\end{equation}
This condition in general does not imply positivity or negativity of the scalar potential and is not invariant under K\"ahler transformations.
In the following we will denote models satisfying eq.~\eqref{weaknoscale} as weakly no-scale. Note also that, if we additionally impose a constant superpotential then eq.~\eqref{weaknoscale} and eq.~\eqref{eq:chiral_no_scale} are equivalent.
\subsection{Linear Multiplets}\label{lin_multi}
In this section we will review $\mathcal{N}=1$ supergravity in four dimensions coupled to $n$ linear multiplets $L^i$. The latter provide a dual description of chiral models with a Peccei-Quinn shift-symmetry. As in the previous section we adopt the notation and formalism of \cite{Binetruy:2000zx, Grimm:2005fa}. Linear multiplets are defined via the constraint equations
\begin{equation}\label{Def_linear_multi}
 (D^2 - 8\bar{R}) L^i = (\bar{D}^2 - 8R) L^i = 0 \ .
\end{equation}
The respective bosonic degrees of freedom are given by $(L^i, B_2^{i})$, where $L^i$ is a real scalar and $B_2^{i}$ a two-form. Contrary to the chiral superfields, the off-shell spectrum of the linear multiplets does not contain any auxiliary field. The geometry of the scalar fields is indirectly determined via two real functions $K(L^i)$ and $F(L^i)$, the former being identical to the K\"ahler potential in the dual chiral formulation.\footnote{More precisely, upon dualizing the theory of linear multiplets to a theory with chiral multiplets, as we discuss in detail in appendix~\ref{duality_matching}, we identify $K(L^i(T))$ as the K\"ahler potential of the chiral theory.} We will refer to it as K\"ahler potential also in theories with linear multiplets. The respective superspace-Lagrangian reads
\begin{equation}\label{Llinear}
 \mathcal{L} = -3 \int E F(L^i) + \frac{1}{2} \int \frac{E}{R} \mathrm{e}^{K/2} W+ \frac{1}{2} \int \frac{E}{\bar{R}} \mathrm{e}^{K/2} \bar{W} \ ,
\end{equation}
where $W$ is a constant superpotential and $F$ is the aforementioned real function. In the Einstein-frame it is implicitly related to the K\"ahler potential as follows
\begin{equation}\label{Einstein_frame}
 1 - \tfrac{1}{3} L^i K_{L^i} = F - L^i F_{L^i} \ ,
\end{equation}
where we abbreviate derivatives with respect to the real scalars as $K_{L^i}$ and $F_{L^i}$ respectively. 
The metric inside the kinetic terms is derived from the so-called kinetic or Hessian potential, which is defined as
\begin{equation}\label{kin_pot}
 \tilde{K}(L^i) = K(L^i) - 3F(L^i) \ .
\end{equation}
The respective metric $\tilde{K}_{L^i L^j}$ is understood as a metric on the underlying real manifold, which is parametrized by the scalars $L^i$. We are now in the position to write down the bosonic component Lagrangian derived from eq.~\eqref{Llinear} \cite{Binetruy:2000zx, Grimm:2005fa}
\begin{equation}\label{L:linear}
  e^{-1} \mathcal{L} = - \tfrac{1}{2} \mathcal{R} + \tfrac{1}{4} \tilde{K}_{L^i L^j} \partial_{\mu} L^i \partial^{\mu} L^j + \tfrac{1}{4} \tilde{K}_{L^i L^j} H_{\mu\nu\rho}^i H^{\mu\nu\rho} {}^j - V(L^i) \ ,
\end{equation}
where $H = d B_2$ denotes the field strength of the two-form and the scalar potential reads
\begin{equation}\label{V:linear}
 V(L^i) = \mathrm{e}^K (L^i K_{L^i} - 3) \lvert W \lvert^2 \ .
\end{equation}
In analogy to the previous section we can define no-scale models via the property
\begin{equation}\label{no-scale:linear}
 L^i K_{L^i} = p \ , \qquad p \in \mathbb{R} \ .
\end{equation}
We immediately observe that any such theory fulfills eq.~\eqref{noscalepot_cases}. 

The dualization of the theory in eq.~\eqref{L:linear} to a chiral theory is explained in appendix~\ref{duality_transformation}. There we reproduce the important steps of this procedure and show the matching of the scalar potentials explicitly.
%
\section{Classification of No-Scale Supergravities}\label{classificationnoscale}
In eq.~\eqref{eq:chiral_no_scale} we introduced the definition of no-scale supergravities for a collection of chiral fields. This definition can be read as a differential equation for the function $G$, which we call no-scale differential equation from now on. Classification of no-scale supergravities, thus, amounts to finding the most general solution to this differential equation and demanding that the resulting theory fulfills all necessary additional requirements. For instance $G$ has to yield a positive-definite K\"ahler metric.\footnote{Here we only determine the solution to the differential equation. To obtain explicit examples, one needs to check the positivity of the K\"ahler metric.} 

Solutions to the general no-scale differential equation for $G$ defined in eq.~\eqref{def_G} are in a one-to-one correspondence with solutions to the homogeneous complex Monge-Amp\`{e}re equation (HCMA). More precisely the HCMA reads
\begin{equation}\label{HCMA}
\det(Y_{T_i \bar{T}_{\bar{\jmath}}})=0  \ .
\end{equation}
The equivalence of the HCMA to the no-scale condition can be stated in the following way: For every solution $Y$ of eq.~\eqref{HCMA} we have a no-scale supergravity with
\begin{equation}\label{defY}
 G=-p \ln Y \ .
\end{equation}
and vice versa. This correspondence is demonstrated explicitly in appendix~\ref{MAeq}. Note that an alternative proof of this equivalence was already given in \cite{Barbieri:1985wq}.

In the following we concentrate on no-scale supergravities with a Peccei-Quinn shift-symmetry. In this case the no-scale differential equation is equivalent to the homogeneous real Monge-Amp\`{e}re equation (HRMA). We derive the general solution to this equation and, thus, provide a classification of the respective no-scale models. A semi-explicit solution to the HRMA for the special case with only three fields (that is $n=3$) was given a long time ago in \cite{Chaundy}.\footnote{The general solution can also be expressed in a different form by means of an analogy with hydrodynamics \cite{Fairlie:1994in}. However, the structure of this solution is fully implicit. Furthermore, the relation to the result of \cite{Chaundy} was investigated in \cite{Fairlie:2011md}. It is also worth noting that the HRMA can be expressed as the Euler-Lagrange equation for a Lagrangian describing a theory of galileons \cite{Deser:2012gm, Fairlie:2011md}.}
It turns out that our result coincides with the solution in \cite{Chaundy} for $n=3$. 

In the second part of this section we determine the general solution to the no-scale condition in eq.~\eqref{no-scale:linear} for a theory with linear multiplets. As a consistency check we explicitly perform the dualization to chiral fields in appendix \ref{duality_matching} and demonstrate that the dual theory matches precisely with the general solution, which we obtained via the HRMA equation.
\subsection{General Solution of the Real Homogeneous Monge-Amp\`{e}re Equation}\label{MAsol}
Let us now turn to the special case, in which the theory possesses a Peccei-Quinn shift-symmetry. More explicitly this means that the theory is invariant under the transformations
\begin{equation}
 T_j - \bar{T}_j \rightarrow T_j - \bar{T}_j + i c_j \ , \qquad j=1,\dots,n  \ ,
\end{equation}
where $c_j \in \mathbb{R}$. Without loss of generality we can assume that the theory is described via a K\"ahler potential of the type $K(T+\bar{T})$ and a constant superpotential.\footnote{By performing an appropriate K\"ahler transformation on a general theory with this shift-symmetry one can always bring the K\"ahler potential and superpotential to this form.} This reduces the no-scale condition to the differential equation for the K\"ahler potential in eq.~\eqref{weaknoscale}, restricting the corresponding K\"ahler manifold $\mathcal{M}$. From now on it is convenient to regard $\mathcal{M}$ as a real Riemannian manifold being parametrized by the real and imaginary parts of $T_i$. Introducing the notation
\begin{equation}\label{def:phi^I}
 \phi_i=\tfrac{1}{2} (T_i + \bar{T}_i) \ ,
\end{equation}
allows to rewrite eq.~\eqref{HCMA} as the HRMA
\begin{equation}\label{HRMA}
 \mathrm{det}(Y_{\phi_i \phi_j}) = 0 \ .
\end{equation}
%
We now deduce the general solution to the HRMA constructively. First, note that eq.~\eqref{HRMA} is equivalent to solving the eigenvalue equation
\begin{equation}\label{eigen}
 Y_{\phi_i \phi_j}(\phi_1,\dots,\phi_n) v_j(\phi_1,\dots,\phi_n)=0 \ , \qquad i=1,\dots, n \ ,
\end{equation}
where $v_j(\phi_1,\dots,\phi_n)$ denotes the respective eigenvector.\footnote{In appendix~\ref{MAeq} we show that $v_j = K^{T_j}$. Here, we introduced a new symbol for the eigenvector for clarity.} 
By defining
\begin{equation}\label{derY}
 Y_{\phi_i} (\phi_1,\dots,\phi_n) = Z^{i}(\phi_1,\dots,\phi_n)  \ ,
\end{equation}
we can rewrite eq.~\eqref{eigen} as 
\begin{equation}\label{FOPDE}
 v_j \partial_{\phi_j} Z^{i}=0 \ .
\end{equation}
Eq.~\eqref{derY} and \eqref{FOPDE} constitute a system of first order homogeneous, linear partial differential equations which is equivalent to eq.~\eqref{eigen}. We proceed by solving this system 
iteratively, that is by integrating eq.~\eqref{FOPDE} first. The latter can be understood geometrically in terms of a vector field $\chi$, given in local (real) coordinates as $\chi = v_j \partial_{\phi_j}$, which annihilates a collection of functions $Z^{1},\dots,Z^n$. 
To this vector field we can associate an integral curve 
\begin{equation}
 \gamma_\chi: I \longrightarrow \mathcal{M} \ , \qquad I \subset \mathbb{R} \ ,
\end{equation}
such that\footnote{Since from appendix~\ref{MAeq} we know that $\chi = K^{T_i} \partial_{\phi_i}$, $\chi$ has to be a nowhere vanishing vector field as demanded by the no-scale condition in eq.~\eqref{weaknoscale}. This guarantees that the integral curve exists in the local patch we are regarding.}
\begin{equation}\label{def_int_curve}
 \frac{\mathrm{d}}{\mathrm{d} s} \gamma_\chi(s) = \chi(\gamma_\chi (s)) \ .
\end{equation}
Eq.~\eqref{FOPDE} implies that the functions $Z^{i}$ are constant along the integral curve $\gamma_\chi$
\begin{equation}\label{dFids}
 \frac{\mathrm{d}}{\mathrm{d} s} Z^{i}(\gamma_\chi(s))=0 \ .
\end{equation}
We can define a new coordinate system 
\begin{equation}
 (\phi_1,\dots,\phi_n) \longrightarrow (u_1(\phi_i),\dots,u_n(\phi_i)) \ , \quad \text{with} \quad  \frac{\partial}{\partial u_n} = \chi \ ,
\end{equation}
that is $u_n$ is a coordinate parametrizing $\gamma_\chi$ and the remaining coordinates $u_\alpha$, $\alpha = 1,\dots,n-1$ are chosen appropriately.\footnote{The imaginary part of $T_i$ are unchanged under this coordinate transformation. Note that, this coordinate transformation does not respect the complex structure of the K\"ahler manifold. However, since the final result is expressed in terms of the original coordinates, the complex structure is restored.} Thus, the solution to eq.~\eqref{dFids} reads
\begin{equation}\label{sol1}
 Z^{i}(\phi_1,\dots,\phi_n) = g^{i}(u_1(\phi_j),\dots ,u_{n-1}(\phi_j)) \ ,
\end{equation}
where $g^i$ is an arbitrary function. 
It remains to integrate eq.~\eqref{derY}. Using the fact that we performed a coordinate transformation and eq.~\eqref{sol1}, we can rewrite eq.~\eqref{derY} as
\begin{equation}
 \partial_{u_j} Y(\phi_1(u_k),\dots,\phi_n(u_k)) = g^{i}(u_1,\cdots ,u_{n-1})\frac{\partial \phi_i}{\partial u_j} \ , \qquad j = 1,\dots,n \ .
\end{equation}
Specifically we have
\begin{equation}
 \partial_{u_n}\left(Y-g^{i}\phi_i\right)=0 \ ,
\end{equation}
which can be integrated directly
\begin{equation}\label{sol2}
 Y(\phi_1,\dots,\phi_n) = \phi_i g^{i}(u_1(\phi_k),\dots ,u_{n-1}(\phi_k))+\tilde{Y}(u_1(\phi_k),\dots ,u_{n-1}(\phi_k)) \ .
\end{equation}
We have to make sure the above integrated form of $Y$ still satisfies eq.~\eqref{derY}.
To this end we compute $Y_{\phi_i}$ 
\begin{equation}\label{Yi=gplus}
 Y_{\phi_i} = g^{i} +  \left( \phi_j  \partial_{u_\alpha} g^{j} + \partial_{u_\alpha}\tilde{Y}\right)\frac{\partial u_{\alpha}}{\partial \phi_i} \ ,
\end{equation}
and, thus, the second term on the r.h.s. of the above equation has to vanish. 
This leads to the supplementary constraint equations 
\begin{equation}\label{sol3}
 \phi_j  \partial_{u_\alpha} g^{j} + \partial_{u_\alpha}\tilde{Y} = 0 \ , \qquad \alpha=1,\dots,n-1 \ .
\end{equation}
In sum, the solution to the HRMA is given by eq.~\eqref{sol2} together with the constraints in eq.~\eqref{sol3}. Since it is necessary to solve the additional constraint equations, the solution is of a semi-explicit form. 
For later purposes we rewrite eq.~\eqref{sol2} as
\begin{equation}\label{sol:chiral}
 Y(\phi_1,\dots,\phi_n) = \phi_i Y_{\phi_i}(\phi_1,\dots,\phi_n)+\tilde{Y}(u_1(\phi_k),\dots ,u_{n-1}(\phi_k)) \ .
\end{equation}

As a final remark, it is worth noting that eq.~\eqref{HRMA} is invariant under the following transformations
\begin{equation}\begin{aligned}\label{affinetrans}
 &Y(\phi_i)  \longrightarrow \lambda Y(\varphi_i) + b^i \phi_i + c \ , \qquad b^i, c, \lambda \in \mathbb{R} \ , \text{where} \\
 &\varphi_i(\phi_j) =  A_{i}^{j} \phi_j + a_i \ , \qquad A \in GL(n) \ , \ a_i \in \mathbb{R} \ .
\end{aligned}\end{equation}
Once a particular solution to the HRMA is identified, one can use the symmetry-transformations to obtain additional solutions. In this sense solutions form equivalence classes under the transformation rules. In particular, we will later on use this to identify seemingly distinct solutions.
%
%
\subsection{Classification of No-Scale Supergravities for Linear Multiplets}\label{classification_linear}
We now turn to the formulation of supergravity coupled to linear multiplets, which we reviewed in sec.~\ref{lin_multi}. 
Recall the definition of a no-scale supergravity in the linear multiplet formalism, which reads
\begin{equation}\label{noscale_linear}
 L^i K_{L^i} = p \ , \qquad p \in \mathbb{R} \ .
\end{equation}
We read this equation as a differential equation for the K\"ahler potential $K$. Contrary to the no-scale differential equation for the chiral multiplets eq.~\eqref{noscale_linear} is of first order. To obtain the correct number of degrees of freedom we need another first order differential equation. The missing equation is the Einstein-frame normalization condition in eq.~\eqref{Einstein_frame}, which relates the function $F$ to the K\"ahler potential. More precisely, inserting eq.~\eqref{noscale_linear} into the Einstein-frame condition we obtain
\begin{equation}\label{noscale_F}
 F - L^i F_{L^i} = 1 - \tfrac{p}{3} \ .
\end{equation}
Eqs.~\eqref{noscale_linear} and \eqref{noscale_F} describe a system of first order differential equations, which specifies no-scale supergravities in the linear multiplet formalism and 
we now solve this system explicitly. 
In the spirit of eq.~\eqref{defY} we introduce
\begin{equation}\label{k=-plogY}
 K(L^i) = - p \,\mathrm{ln}(Y(L^i)) \ ,
\end{equation}
which when inserted in eq.~\eqref{noscale_linear} yields
\begin{equation}\label{pdeY}
 L^i Y_{L^i} = - Y \ .
\end{equation}
Note that a homogeneous function $\mathcal{F}(\phi_1,\dots,\phi_n)$ of degree $m$ is defined via the property
\begin{equation}\label{def_hom}
 \mathcal{F}(\lambda \phi_1,\dots,\lambda \phi_n) = \lambda^m \mathcal{F}(\phi_1,\dots,\phi_n) \ , \qquad \forall \lambda \in \mathbb{R} \ .
\end{equation}
Alternatively homogeneous functions can be defined as the general solution to the following differential equation
\begin{equation}
 m \mathcal{F} = \phi_i \mathcal{F}_{\phi_i} \ .
\end{equation}
Thus, the general solution to eq.~\eqref{pdeY} is a homogeneous function of degree $-1$. In particular eq.~\eqref{def_hom} implies that $Y$ satisfies\footnote{Strictly speaking the identification is given as $Y(L^1,\dots,L^n) = \tfrac{1}{L^1} Y (1,\tfrac{L^2}{L^1},\dots,\tfrac{L^n}{L^1})$.}
\begin{equation}\label{sol_lin_1}
 Y(L^1,\dots,L^n) = \frac{1}{L^1} Y\left(\frac{L^2}{L^1},\dots,\frac{L^n}{L^1} \right) \ .
\end{equation}
It remains to solve eq.~\eqref{noscale_F}, which is an inhomogeneous linear differential equation. The general solution can be written as the sum of the general solution to the respective homogeneous equation and a particular solution to the inhomogeneous equation. The corresponding homogeneous equation is the differential equation for a homogeneous function of degree one, so that we can write the solution as
\begin{equation}\label{sol_lin_2}
 F(L^i) = F_{(1)}(L^i) + 1 - \tfrac{p}{3} \ ,
\end{equation}
where $F_{(1)}$ is homogeneous of degree one. As promised we find that the solution to the system of eqs.~\eqref{noscale_linear} and \eqref{noscale_F} can be displayed explicitly.
\section{Explicit Classes of Solutions and Remarks on Geometry}\label{explicitsolutions}
In the previous section we derived the general solution to the no-scale condition both for a theory of chiral multiplets with a shift-symmetry as well as for a theory with linear multiplets. Furthermore, in appendix~\ref{duality_matching} we show that these solutions match upon dualization. The result for the chiral theory, which is displayed in eq.~\eqref{sol2} and \eqref{sol3}, is semi-explicit. To find fully explicit solutions we need to impose additional conditions. It will be the purpose of this section to construct classes of explicit solutions and make a comparison to those examples, which are already known in the literature. 

In the second part of this section we will derive certain geometric statements, which on one hand illustrate the geometric meaning of the no-scale condition, and on the other hand help to understand the difference between the classes of solutions which we present in the first part of this section.

Before turning to the examples, let us make a few comments on how to interpret the solution in eq.~\eqref{sol2} and \eqref{sol3}. One may regard eqs.~\eqref{sol3} as a system of differential equations, which needs to be solved in order to find explicit solutions. The input data for the initial value problem corresponding to eq.~\eqref{HRMA} are two functions, that are allowed to depend on $(n-1)$ variables each. The standard procedure (although complicated) is to choose two functions out of the collection $(g^{1}, \dots, g^{n}, \tilde{Y})$ and determine the remaining functions by solving the system in eqs.~\eqref{sol3}. In this line of reasoning the functions $u_\alpha (\phi)$ are related to the coordinates with respect to which we define a hyper-surface, on which we specify the initial data. For instance we could choose the initial data on a hyper surface defined via the equation $u_n = 0$. 
For the practical purpose of finding explicit solutions it turns out to be advantageous to simply choose all $(g^{1}, \dots, g^{n},  \tilde{Y})$ and interpret eqs.~\eqref{sol3} algebraically, that is to solve for $u_\alpha$. As long as the system is solvable this procedure automatically yields a solution. As we will see it is in fact not always necessary to choose all $(g^1, \dots, g^n,  \tilde{Y})$, but possibly only a subclass.
\subsection{Homogeneous Functions}\label{hom_functions}
We now apply the aforementioned procedure to construct a class of explicit solutions for the chiral theory. Let us choose 
\begin{equation}\label{ktildehom}
 \tilde{Y}=0 \ ,
\end{equation}
while keeping $g^1,\dots,g^n$ arbitrary. Eq.~\eqref{sol:chiral} then reduces to
\begin{equation}\label{homsol}
 Y(\phi_1,\dots, \phi_n) =  \phi_i Y_{\phi_i}(\phi_1,\dots, \phi_n) \ .
\end{equation}
In eq.~\eqref{homsol} we recognize the differential equation, which defines a homogeneous function of degree one.\footnote{Recall that solutions form equivalence classes, where individual representatives are related by the symmetries given in eq.~\eqref{affinetrans}. In particular a homogeneous function in general transforms into an inhomogeneous one.}
Let us also display the form of the solutions satisfying eq.~\eqref{ktildehom} in the dual linear multiplet formulation, see appendix \ref{duality_matching} for the relations between the solutions in both formalisms.
Using eqs.~\eqref{ktildehom}, \eqref{k_linear} and \eqref{sol_lin_2} we obtain that the homogeneous solutions correspond to a constant $F$ given by
\begin{equation}\label{Fhom}
  F=1-\frac{p}{3} \ .
\end{equation}
Furthermore, recall that the HRMA is equivalent to the existence of an eigenvector with eigenvalue zero. In appendix~\ref{MAeq} we determine this eigenvector explicitly. For the class of homogeneous functions it turns out that the components of this eigenvector are particularly simple. Namely, taking the derivative of eq.~\eqref{homsol} yields the looked-for eigenvalue equation~\eqref{eigen} with corresponding eigenvector
\begin{equation}
 v_j = \phi_j\ .
\end{equation}
Similarly direct computation yields
\begin{equation}\label{K^T=phi}
 K^{T_i} = - 2 \phi_i \ , 
\end{equation}
which is in agreement with the identification of the eigenvector according to appendix~\ref{MAeq}.

Logarithmic K\"ahler potentials which satisfy eq.~\eqref{homsol} are important in the context of moduli spaces of superstring compactifications. A well known example is the large-volume limit of the K\"ahler potentials of the moduli spaces of type IIB Calabi-Yau orientifold compactifications with $O3/O7$-planes \cite{Giddings:2001yu, Becker:2002nn, Grimm:2004uq}. Here the K\"ahler moduli sector has a perturbative Peccei-Quinn shift symmetry and, thus, the superpotential is constant along the K\"ahler moduli directions. The respective K\"{a}hler potential here is of the form
\begin{equation}\label{LVtypeII}
K(T,\bar{T}) = - 2 \ln\mathcal{V}(T,\bar{T})\ ,\quad \mathcal{V}(T,\bar{T}) = \mathcal{K}_{ijk} t^i t^j t^k 
\end{equation}
where $\mathcal{K}_{ijk}$ are the triple intersection numbers and $t^i$ are two-cycle volumes. The latter are related to the K\"{a}hler moduli through $\text{Re}T_i = \frac{3}{2}\mathcal{K}_{ijk}t^j t^k$. From \eqref{LVtypeII} we learn that $\mathcal{V}$ is homogeneous of degree $3$ in $t^i$ which, in turn implies it is of degree $3/2$ in $\text{Re}T_i$.\footnote{Notice that here $Y = \mathcal{V}^{\frac{2}{3}}$ and, thus, $Y$ is a homogeneous function of degree one.}  
Automatically this leads to the no-scale property
\begin{equation} \label{noscaleLVtzpeII}
G^{T_i \bar{T}_j} G_{T_i} G_{\bar{T}_j} = 3 \ . 
\end{equation}
Moreover, let us display the K\"ahler potential in terms of linear multiplets. Using eq.~\eqref{Fhom} we find that $F=0$, since $p=3$ by means of eq.~\eqref{noscaleLVtzpeII}.
Using eq.~\eqref{F+TL=1} we determine $L^i$ to be
\begin{equation}
L^i = \frac{3}{2} \frac{t^i}{\mathcal{V}} \ . 
\end{equation}
We are now in a position to express the K\"{a}hler potential as a function of the real linear scalars 
\begin{equation}
K(L) = \ln(\tfrac{8}{27}\mathcal{K}_{ijk} L^i L^j L^k) \ , 
\end{equation}
which as expected is a homogeneous function of degree $3$.\footnote{Analogously to the previous footnote, in this example $Y(L) = (\tfrac{8}{27}\mathcal{K}_{ijk} L^i L^j L^k)^{-2/3}$ and, hence, $Y(L)$ is an homogeneous function of degree $-1$.} We will return to the discussion of moduli spaces of superstring compactifications in sec.~\ref{noscale_string}, where we will make a broader connection between logarithmic K\"ahler potentials of a homogeneous type and the geometry of moduli spaces.
%
\subsection{Non-Homogeneous Examples}\label{sec:nonhom}
In the previous section we have displayed an explicit class of solutions to eq.~\eqref{HRMA}, namely the homogeneous functions and furthermore hinted on their importance in the context of compactifications of string theory. In this section we want to illustrate explicitly that the general solution in eq.~\eqref{sol2} encompasses more than just homogeneous functions. Even though these non-homogeneous solutions might look involved they provide new insights into the expected K\"{a}hler potentials.

We begin by making a general observation. Given a particular solution $Y_0(\phi_1,\dots,\phi_n)$ to the HRMA with $n$ variables $\phi_1,\dots,\phi_n$ one can construct a solution $Y$ to the HRMA with $n+m$ variables $M_I = (\phi_i, \rho_a)$, where $i = 1,\dots,n $ and $a = 1, \dots, m$ via
\begin{equation}\label{new_sol_Y}
 Y(M_I) = Y_0(\phi_1 + Q_1(\rho_a), \dots, \phi_n + Q_n(\rho_a)) \ , 
\end{equation}
with $Q_1,\dots,Q_n$ being a collection of real-valued functions. Direct computation yields that $Y$ in eq.~\eqref{new_sol_Y} obeys
\begin{equation}
 Y_{M_I M_J} v_{J} = 0 \ , 
\end{equation}
where $v_J = (v_i^{(0)},0)$ and $v^{(0)}$ denotes the eigenvector to eigenvalue zero associated with the fact that $Y_0$ solves the HRMA. For the simple case $n=1$ the above construction leads to a solution of the form
\begin{equation}
 Y = \phi_1 + Q_1 (\rho_1,\dots,\rho_m)\ ,
\end{equation}
which for general $Q_1$ is non-homogeneous. Note that, the linear term in the above equation can also be absorbed by performing a symmetry-transformation of the type in eq.~\eqref{affinetrans}. Thereby the function transforms into another function which depends only on $\rho_1, \dots, \rho_m$ and, thus, trivially solves the HRMA. Furthermore, one may consider $Y_0$ to be a homogeneous function of degree one, which for general $Q_i$ again leads to non-homogeneous examples. However, note that solutions of the type in eq.~\eqref{new_sol_Y} are very special, since they effectively only depend on $n$ variables. 

Thus, it is interesting to construct solutions which fully exemplify the generality of eq.~\eqref{sol1}. To this end we need to solve the constraint equations in eq.~\eqref{sol3} which in general is a delicate task. In what follows we construct solutions to eq.~\eqref{sol3} for the special case of two fields, that is $n=2$. The respective K\"ahler potentials were so far unknown in the literature.

For a two-dimensional K\"ahler manifold there is only one coordinate $u$ and one constraint equation. The latter can be recast into the form
\begin{equation}\label{ce_2d}
  \phi_1 + \phi_2 \frac{g^{2\,\prime}}{g^{1\,\prime}} + \frac{\tilde{Y}^\prime}{g^{1\,\prime}} = 0 \ ,
\end{equation}
provided $g^{1\,\prime}\neq0$ and where $g^1,g^2$ and $\tilde{Y}$ are functions of $u$.
We define $f \equiv g^{2\,\prime} / g^{1\,\prime}$ and assume $\tilde{Y}^\prime /g^{1\,\prime} = f^q$ with $q\in \mathbb{Z}$, while keeping $g^1$ and $f$ arbitrary. 
Eq.~\eqref{ce_2d} can be analytically solved in very few cases. Namely for $q=1,2,3,4$ the roots of eq.~\eqref{ce_2d} and, thus, $u$ can be determined explicitly. For $q=1$ the solution can be transformed into a homogeneous function by means of eq.~\eqref{affinetrans}. For $q=2$, there are two solutions for $u$ given by \footnote{To ensure that $u$ is a real function we have to restrict the field range to the region $\phi_2^2 \geq 4 \phi_1$.}
\begin{equation}\label{sol_u}
 u_{\pm}(\phi_1,\phi_2) = f^{-1}\left(-\tfrac{1}{2} \phi_2 \pm \sqrt{\tfrac{1}{4} \phi_2^2-\phi_1} \right) \ .
\end{equation}
For $q=3,4$ the solutions are quite complicated and we will not display them here. In general expressing $g^2$ and $\tilde{Y}$ in terms of $g^1$ and $f$, the solution reads
\begin{equation}\begin{aligned}
   Y(\phi_1,\phi_2) &= [\phi_1 + \phi_2 f(u) + f^q(u)] \, g^1(u)  \\
   & \quad - \phi_2 \int \mathrm{d}u \,g^1(u) f^{\prime}(u) - \int \mathrm{d}u \, g^1 (u) \, (f^q)^{\prime}(u) \ ,
\end{aligned}\end{equation}
where $u$ is understood as being replaced by the solution $u(\phi_1,\phi_2)$ to eq.~\eqref{ce_2d}. Let us now look for simple particular examples. Setting $q=2$ and $g^1(u) = f(u) = u$ the solution reduces to 
\begin{equation}\label{eq:uu}
Y(\phi_1, \phi_2) = \frac{1}{12}\left[ \phi_2^3 - 6 \phi_1 \phi_2 + (\phi_2^2 - 4\phi_1)^{3/2}  \right] \ .
\end{equation}
Another interesting example is $q=2$ and $g^1(u) = u, f(u) = \ln u$, where \footnote{We chose both for eq.~\eqref{eq:uu} as well as for $Y$ below the solution $u_-$, for which we explicitly checked that the respective $Y$ yield a real valued K\"ahler potential and a positive-definite K\"ahler metric in a certain region of field space.}
\begin{equation}\label{nolog}
Y(\phi_1, \phi_2) = \exp\left[-\frac{1}{2}\left(\phi_2 - \sqrt{\phi_2^2 - 4\phi_1} \right)\right] \left(2 - \sqrt{\phi_2^2-4\phi_1}\right) \ .
\end{equation}
Notice here that after replacing $Y$ in the K\"{a}hler potential, as defined in eq.~\eqref{defY}, the first factor drops out of the logarithm. It is somewhat surprising that a K\"{a}hler potential of this form obeys the no-scale condition.\footnote{Note also, that by means of a K\"ahler transformation we can recast the theory in eq.~\eqref{nolog} as a theory with $W= \mathrm{exp}(\tfrac{p}{4}T_2)$ and $K = -p\ln(2 - \Delta)-\tfrac{p}{2} \Delta$ where $\Delta(\phi_1,\phi_2) = \sqrt{\phi_2^2 - 4\phi_1}$.}
\subsection{Remarks about Curvature and Geometry}
After derivation of the general solution to the no-scale condition for models with shift-symmetry it is interesting to make statements about the resulting geometries. We will not attempt to compute the components of the Riemann-tensor here, but try to offer some general insights. Firstly, we will derive a statement about the holomorphic sectional curvature for general no-scale models. Secondly, we will derive a geometric distinction between homogeneous and non-homogeneous solutions. 

In \cite{Covi:2008ea} it was shown that no-scale models have constant holomorphic sectional curvature along a special direction. Let us here slightly generalize the derivation of \cite{Covi:2008ea} starting from the general no-scale condition in eq.~\eqref{eq:chiral_no_scale}. In the following we denote the covariant derivative on the K\"ahler manifold by $\nabla_{T_i}$. This covariant derivative is defined with respect to a metric compatible and hermitian connection. To begin with, taking the covariant derivative of eq.~\eqref{eq:chiral_no_scale} we deduce
\begin{equation}\label{del1}
 G_{T_i} + G^{T_k} \nabla_{T_i} G_{T_k} = 0 \ ,
\end{equation}
where we used the fact that $\nabla_{T_i} G_{\bar{T}_{\bar{k}}} = G_{T_i \bar{T}_{\bar{k}}}$ as well as metric compatibility. Taking an anti-holomorphic covariant derivative of eq.~\eqref{del1} one infers
\begin{equation}\label{del2}
 G_{T_i \bar{T}_{\bar{\jmath}}} + \nabla_{T_i} G_{T_k} \nabla_{ \bar{T}_{\bar{\jmath}}} G^{T_k} - R_{T_i \bar{T}_{\bar{\jmath}} T_k \bar{T}_{\bar{l}}} G^{T_k} G^{\bar{T}_{\bar{l}}} = 0 \ ,
\end{equation}
where $R$ denotes the Riemann tensor. Instead taking a holomorphic derivative of eq.~\eqref{del1} yields
\begin{equation}\label{del3}
 2 \nabla_{T_i} G_{T_j} + G_{T_k} \nabla_{T_i} \nabla_{T_j} G^{T_k} = 0 \ .
\end{equation}
Using eq.~\eqref{del1} and eq.~\eqref{del3} we can derive the auxiliary property
\begin{equation}
  \nabla_{T_i} G_{T_k} \nabla_{ \bar{T}_{\bar{\jmath}}} G^{T_k} =  \nabla_{T_k} G_{T_i} \nabla_{ \bar{T}_{\bar{\jmath}}} G^{T_k} \ .
\end{equation}
Altogether and using the no-scale condition in eq.~\eqref{eq:chiral_no_scale} we find that
\begin{equation}
 R_{T_i \bar{T}_{\bar{\jmath}} T_k \bar{T}_{\bar{l}}} G^{T_i} G^{\bar{T}_{\bar{\jmath}}} G^{T_k} G^{\bar{T}_{\bar{l}}} = 2p \ ,
\end{equation}
which states that all no-scale models have constant holomorphic sectional curvature along the vector field $G^{T_i}$. Recall from appendix~\ref{MAeq} that this vector field is coincident with the eigenvector (field) to the eigenvalue zero ensuring the validity of the HCMA. More specifically in the shift-symmetric case this tells us that the curvature along the integral curve $\gamma_D$ as defined in eq.~\eqref{def_int_curve} is constant.

Let us now turn to shift-symmetric no-scale models and try to make a geometric distinction between K\"ahler potentials of homogeneous and non-homogeneous type. Firstly, for the former it turns out that also the Ricci curvature is constant along $G^{T_k}$. More precisely contracting eq.~\eqref{del2} with an inverse metric and using eq.~\eqref{K^T=phi} we find
\begin{equation}\label{Ric}
  Ric_{T_i \bar{T}_{\bar{\jmath}} }G^{T_i} G^{\bar{T}_{\bar{\jmath}}} = 2n \ ,
\end{equation}
where $Ric$ denotes the Ricci tensor.\footnote{This was also found in \cite{Ferrara:1994kg}.} For non-homogeneous solutions one finds that the l.h.s. of eq.~\eqref{Ric} is in general non-constant. We checked this explicitly for the examples discussed in section \ref{sec:nonhom}.

Moreover, homogeneous-type K\"ahler potentials have an additional isometry. The respective transformations are dilatations, whose infinitesimal version reads 
\begin{equation}
 \delta T_i = \epsilon T_i \ , \qquad \delta \bar{T}_{\bar{\jmath}} = \epsilon \bar{T}_{\bar{\jmath}} \ .
\end{equation}
We identify the respective Killing vector as
\begin{equation}\label{Killing_vector}
 \Xi = \Xi^{T_i} \frac{\partial}{\partial T_i} + \Xi^{\bar{T}_{\bar{\jmath}}} \frac{\partial}{\partial \bar{T}_{\bar{\jmath}}} \ , \qquad \text{where} \qquad \Xi^{T_i} = T_i  \ , \qquad \Xi^{\bar{T}_{\bar{\jmath}}} = \bar{T}_{\bar{\jmath}} \ .
\end{equation}
Indeed, by direct computation (and using eq.~\eqref{K^T=phi}) we find
\begin{equation}\begin{aligned}
 &\nabla_{T_i} \Xi_{T_j} + \nabla_{T_j} \Xi_{T_i} = 0 \ , \\
 &\nabla_{T_i} \Xi_{\bar{T}_{\bar{\jmath}}} + \nabla_{\bar{T}_{\bar{\jmath}}} \Xi_{T_i} = 0 \ ,
\end{aligned}\end{equation}
which are the Killing vector equations.

We can promote this isometry to a symmetry of the bosonic sector of the respective supergravity in eq.~\eqref{L:chiral} if $V=0$, that is if $p=3$, and if the spacetime-metric does not transform under dilatations. For $p$ arbitrary another interesting situation occurs when the spacetime-metric transforms under dilatations with weight $p$. The respective infinitesimal version reads
\begin{equation}
 \delta g_{\mu \nu} = \epsilon p g_{\mu\nu} \ .
\end{equation}
In this case also the bosonic Lagrangian transforms under the dilatations as
\begin{equation}
 \delta \mathcal{L} = \epsilon p \mathcal{L} \ .
\end{equation}
In the context of effective supergravities of string compactifications the scaling property of the Lagrangian plays a role and we will come back to it in the next section. 
\section{No-Scale K\"{a}hler Potentials from String Theory}\label{noscale_string}
In the previous section we presented an explicit example of the general solution to the no-scale condition in sec.~\eqref{classificationnoscale} both for chiral and linear multiplets, namely 
type IIB orientifold flux-compactifications with $O3$-planes. In this section we will study the structure of no-scale models descending from string-theory compactifications in more detail. In general, string-derived no-scale models exhibit a shift-symmetry only in a subsector of the theory and, thus, our setup applies to those cases, in which the total K\"ahler manifold $\mathcal{M}$ of the theory admits a product structure
\begin{equation}\label{strong_no_scale}
 \mathcal{M} = \mathcal{M}_{ns} \times \mathcal{M}_r \ ,
\end{equation}
where $\mathcal{M}_{ns}$ is the subspace of scalars with a shift-symmetry obeying the no-scale property and $\mathcal{M}_r$ is parametrized by the remaining fields. To guarantee the no-scale property in eq.~\eqref{eq:chiral_no_scale} it is also necessary that the superpotential depends only on the fields inside $\mathcal{M}_r$. However, in the context of geometric string compactifications this structure is rather special and appears essentially only in an asymptotic regime of the manifold or for special configurations of the compactification data.\footnote{It is for instance possible that special flux choices in type II orientifold compactifications imply a shift-symmetry in the superpotential. Furthermore, eq.~\eqref{strong_no_scale} holds in type IIB orientifold flux-compactifications with $O3$-planes but no $O7$-planes.} A particular example is the large volume limit of the heterotic string compactified on a Calabi-Yau threefold \cite{Strominger:1985ks, Dixon:1989fj, Candelas:1990pi}. In this example the K\"ahler potential at lowest order in the matter fields is given by a logarithm of a homogeneous function as already pointed out in \cite{Ferrara:1994kg}.\footnote{Note furthermore that terms quadratic in the matter fields also have a homogeneity property as pointed out in \cite{Ferrara:1994kg}.}

It is interesting to turn to the more general situation, in which the theory has a shift-symmetric K\"ahler potential but not necessarily a shift-symmetric superpotential. This allows to discuss the weak-type no-scale condition in eq.~\eqref{weaknoscale}. 
In the following we revisit orientifold compactifications of type IIA/B string theory. These yield 4D theories with a shift-symmetric K\"ahler potential obeying a weak-type no-scale condition at tree level.\footnote{Also in Calabi-Yau fourfold-compactifications of M-theory with flux one can find weak-type no-scale conditions where the respective K\"ahler potential has a homogeneity property \cite{Haack:2001jz}.} Moreover, this condition holds on the entire scalar manifold and not only at special points, such as e.g.~at large volume. As it turns out 
the respective K\"ahler potentials are again given by logarithms of homogeneous functions. 
Furthermore, we will present an example where both the shift-symmetry as well as the weak-type no-scale condition survive perturbative $\alpha'$-corrections.
\subsection{Type IIA/B Orientifold Flux-Compactifications}
The low-energy effective supergravities describing type IIB orientifold-compactifications with fluxes and $O3/O7$-planes or $O5/O9$-planes were determined in \cite{Grimm:2004uq, Grimm:2004ua}. The respective K\"ahler moduli/dilaton subsectors of the effective actions enjoy a shift-symmetric K\"ahler potential, which is given as a logarithm of a homogeneous function of degree $-4$ as stated in \cite{Grimm:2005fa}.\footnote{This structure appears also in generalized orientifold compactifications, for an overview see for instance \cite{Gurrieri:2002wz, Grana:2005ny, Benmachiche:2006df, Grana:2006hr}.} For illustrative purposes we now revisit the $O3/O7$-plane case and demonstrate the homogeneity explicitly. The relevant K\"ahler variables here are given by $T_i, \tau, G_a$, where $\alpha = 1, \dots, h_+^{1,1}$ and $a = 1, \dots,h_-^{1,1}$ and $h^{1,1}_{\pm}$ denote the dimension of the even/odd spaces of harmonic $(1,1)$-forms on the Calabi-Yau orientifold. The K\"ahler potential reads \cite{Grimm:2004uq}
\begin{equation}\label{IIBK}
 K(\tau,\bar{\tau},T,\bar{T},G,\bar{G}) = - \ln(-i(\tau-\bar{\tau})) -2 \ln (\tfrac{1}{6}\mathcal{K}_{ijk} t^i t^j t^k)
\end{equation}
where $t^i$ denote two-cycle volumes as in eq.~\eqref{LVtypeII} and are understood as being replaced by the solution to the equation
\begin{equation}\label{orientifolds}
 T_i + \bar{T}_i  = \frac{3}{2} \mathcal{K}_{ijk} t^j t^k + \frac{3i}{4} \frac{1}{\tau-\bar{\tau}} \mathcal{K}_{i b c} (G_b-\bar{G}_b) (G_c-\bar{G}_c)
\end{equation}
and, thus, implicitly depend on $M_I = (T_i,\tau,G_a)$. Here, the objects $ \mathcal{K}_{ijk}$ and $\mathcal{K}_{i b c}$ denote triple intersection numbers of the compactification manifold. Immediately we observe that $t^i$ and, hence, also the K\"ahler potential has a shift-symmetry with respect to $\mathrm{Im}(T_i)$, $\mathrm{Re}(\tau)$ and $\mathrm{Re}(G_a)$. Moreover, eq.~\eqref{orientifolds} implies that $t^i$ is a homogeneous function of degree $1/2$ in the variables $M_I$. Thus, the K\"ahler potential is given as the logarithm of a homogeneous function of degree $-4$. This in turn yields the following no-scale property
\begin{equation}\label{orientifold_noscale}
 K_{M_I} K^{M_I \bar{M}_J} K_{\bar{M}_J} = 4 \ .
\end{equation}
Note that we can also understand eq.~\eqref{orientifolds} as a particular realization of the construction in eq.~\eqref{new_sol_Y} by identifying $\rho_a = -i(G_a-\bar{G}_a)/2$ and $Y_0$ via eq.~\eqref{LVtypeII}.\footnote{In particular this construction can be used to infer that $K_{m_I} K^{m_I \bar{m}_J} K_{\bar{m}_J} = 3$ where $m_I = (T_i,G_a)$, that is by setting the dilaton to a constant.}

Let us remark that since here the superpotential depends on the dilaton \cite{Grimm:2004uq}, the weak no-scale condition alone does not imply positivity of the scalar potential. Nevertheless, it turns out that the scalar potential is positive-semi-definite for eq.~\eqref{IIBK}. This can be well understood in a dual description where $T_i$ are dualized into linear multiplets \cite{Grimm:2004uq}.\footnote{Recall from our discussion in the preceding sections, that for models with a Peccei-Quinn shift-symmetry the no-scale conditions in the chiral theory and in the dual theory with linear multiplets are equivalent to each other. We would like to emphasize here, that this equivalence in general does not hold any longer, if the theory includes additional chiral fields, which do not exhibit a Peccei-Quinn shift-symmetry and hence can not be dualized to linear multiplets.}

The effective action of type IIA orientifold flux-compactifications with $O6$-planes was computed in \cite{Grimm:2004ua}. Here, the K\"ahler potential has a shift-symmetry in the real (or imaginary) parts of complex structure, dilaton and K\"ahler moduli and, furthermore, is given as a logarithm of a homogeneous function of degree $-7$. Thus, one has a weak no-scale property
\begin{equation}
  K_{\mathcal{E}_A}K^{\mathcal{E}_A \bar{\mathcal{E}}_{\bar{B}}}K_{\bar{\mathcal{E}}_{\bar{B}}} = 7 \ ,
\end{equation}
where $\mathcal{E}_A$ collectively denote all complex scalar fields in the spectrum.
\subsection{Type IIB Flux-Compactifications with $O3$-Planes and $(\alpha')^3$-Corrections}
The 10D effective action of type IIB string theory receives perturbative $\alpha'$- and $g_s$-corrections. In particular this includes a term at order $(\alpha')^3$ consisting of contractions of four Riemann-tensors \cite{Antoniadis:1997eg}. This term can be understood as a four-point scattering amplitude, receiving contributions at string tree-level as well as at string-loop level. For Calabi-Yau orientifold flux-compactifications with $O3$-planes a correction to the K\"ahler potential of the K\"ahler moduli is induced by this 10D correction and was determined in \cite{Becker:2002nn}. At string tree-level the $\alpha'$-corrected K\"ahler potential for the dilaton/K\"ahler-moduli sector reads
\begin{equation}\label{K_alphaprime}
 K(\tau,\bar{\tau},T_i,\bar{T}_{\bar{\jmath}}) = - \ln(-i(\tau-\bar{\tau})) -2 \ln \left(\mathcal{V}(T_i,\bar{T}_{\bar{\jmath}}) + \xi [-\tfrac{i}{2}(\tau-\bar{\tau}) ]^{3/2} \right) \ .
\end{equation}
Here $\xi$ is a numerical constant and the K\"ahler variables $T_i$ are related to the volume $\mathcal{V}$ in the same way as in eq.~\eqref{LVtypeII}. Again we have a shift-symmetry both for $T_i$ as well as for the dilaton. Moreover, from sec.~\ref{hom_functions} we know that $\mathcal{V}$ is a homogeneous function of degree $3/2$ in $T_i + \bar{T}_i$. Hence, taking into account the dilaton-dependence we find that $K$ in eq.~\eqref{K_alphaprime} is given by a logarithm of a homogeneous function of degree $-4$. This yields the following no-scale property
\begin{equation}\label{no_scale_alphaprime}
 K_{N_A} K^{N_A \bar{N}_B} K_{\bar{N}_B} = 4 \ ,
\end{equation}
where $N_A = (T_i,\tau)$.\footnote{There is also a notable example, where perturbative $\alpha$-corrections preserve the no-scale structure \cite{Grimm:2013bha}. More precisely these are induced by the leading order $\alpha$-corrections to the 11d low-energy effective action of M-theory after compactification on a Calabi-Yau fourfold. Furthermore, in the correct K\"ahler coordinates the K\"ahler potential has a homogeneity property. This is not immediately visible in the chiral description, but can be more easily inferred from the dual description in terms of linear multiplets given in that reference.} 
To our knowledge eq.~\eqref{no_scale_alphaprime} was so far unknown. Let us emphasize again that this is a weak-type no-scale property. Contrary to the previous example, here the scalar potential is not positive semi-definite \cite{Becker:2002nn}. Moreover, it is interesting to note that eq.~\eqref{no_scale_alphaprime} holds no longer after including string-loop corrections to the 10D higher-curvature invariant. It was found in \cite{Antoniadis:1997eg} that these corrections essentially coincide with the tree-level $R^4$-term, the only difference being that they are suppressed with an additional factor of $g_s^2$. 
In turn, we find that the K\"ahler potential for the effective 4D action of the K\"ahler moduli-dilaton subsector after including the one-loop corrections is of the form
\begin{equation}\begin{aligned} \label{IIB_with_gs}
K(\tau,\bar{\tau},T_i,\bar{T}_{\bar{\jmath}}) &= - \ln(-i(\tau-\bar{\tau})) \\
& \quad -2 \ln \left(\mathcal{V} + \xi [-\tfrac{i}{2}(\tau-\bar{\tau})]^{3/2} + \tilde{\xi} [-\tfrac{i}{2}(\tau-\bar{\tau})]^{-1/2}\right) \ ,
\end{aligned}\end{equation}
where $\tilde{\xi}$ is another numerical factor. We immediately observe that the homogeneity of the K\"ahler potential is spoiled.
\subsection{Discussion}
One might wonder whether it is possible to understand the form of the K\"ahler potentials of type II compactifications with orientifolds and in particular the weak no-scale property more conceptually. Some insight can be gained from the scale invariance of string amplitudes at tree level. As argued in \cite{Witten:1985xb} for the heterotic string, this scale invariance induces a scaling property of the 10D effective Lagrangian. This in turn implies that the 4D effective Lagrangian obtained after compactification transforms with a certain weight under dilatations. However, in this case the only fields in the 4D theory that transform non-trivially are the dilaton and the space-time metric. In this way the scaling property of the Lagrangian does not constrain the form of the K\"ahler potential of the K\"ahler moduli. On the other hand for orientifold compactifications of the type II string we expect that the 4D fields will transform with different weights and so in principle the scaling property of the Lagrangian can lead to conditions upon the K\"ahler potential of the K\"ahler moduli.  
Recall that logarithmic K\"ahler potentials of homogeneous type possess an isometry associated with dilatations, where the respective Killing vector is given in eq.~\eqref{Killing_vector}. It is conceivable that this isometry is induced by the scaling behaviour of the Lagrangian. Moreover, the scale-invariance of string amplitudes is explicitly broken at the quantum level and, hence, does not survive $g_s$-corrections \cite{Witten:1985xb}. This fact is in agreement with the observation that the $\alpha'$-corrected K\"ahler potential in eq.~\eqref{K_alphaprime} is still given by a logarithm of a homogeneous function while the $g_s$-corrected K\"ahler potential in eq.~\eqref{IIB_with_gs} is not.

One might go even further and ask whether the scaling behaviour is responsible for the weak no-scale property. However, the following evidence suggests otherwise. 
Consider, for example, that fluxes are absent so that the superpotential in the effective 4D theory vanishes identically. In this case only the scaling behavior of the kinetic terms constrains the form of the K\"ahler potential. Given that the fields in the spectrum transform with certain weights under dilatations one finds that a K\"ahler potential which is a homogeneous function of an appropriate degree will support the required weight of the kinetic terms.
However, no such K\"ahler potential is of the weak no-scale type, which can be checked explicitly. 

In conclusion, it is evident that the scaling behaviour of the Lagrangian implies the existence of the Killing vector in eq.~\eqref{Killing_vector}, but is not strong enough to enforce the weak no-scale property and, hence, to uniquely single out logarithmic K\"ahler potentials of homogeneous type. Seen from this perspective, it could be possible that the weak no-scale property holds also at the quantum level, although our knowledge about the effective 4D action is far too incomplete to verify this yet. Moreover, since the scaling symmetry is broken at loop-order, any K\"ahler potential that supports the weak no-scale property beyond tree-level is very likely going to be of a non-homogeneous type. 

Furthermore, it would be interesting to explore whether stringy 4D effective Lagrangians exist, which do not inherit a scaling property from string amplitudes and, thus, may constitute examples of non-homogeneous no-scale models. One may for instance wonder whether such models could arise from non-geometric string compactifications.
\section*{Acknowledgements}
We would like to thank Ido Ben-Dayan, Jan Louis and Severin L\"{u}st for important comments and discussions. We also thank Peter-Simon Dieterich, Klaus Fredenhagen, Constantin Muranaka and Alexander Westphal. This work was supported by the German Science Foundation (DFG) under the Collaborative Research Center (SFB) 676 Particles, Strings and the Early
Universe, and by the Impuls und Vernetzungsfond of the Helmholtz Association 
of German Research Centers under grant HZ-NG-603.
%
%
\appendix
\section{Equivalence to the Homogeneous Monge-Amp\`{e}re Equation}\label{MAeq}
In this appendix we explicitly demonstrate the equivalence between the no-scale condition for chiral fields, as given in eq.~\eqref{eq:chiral_no_scale} and the HCMA, see eq.~\eqref{HCMA}. Moreover, this demonstration will provide additional insight into the geometric meaning of the no-scale condition. Firstly, the HMCA, given in eq.~\eqref{HCMA}, is equivalent to the statement that $Y_{T_i \bar{T}_{\bar{\jmath}}}$ has an eigenvalue zero. For $G$ given in eq.~\eqref{defY} we will now show that the respective eigenvector is given by 
\begin{equation}
 G^{T_i} \equiv G^{T_i  \bar{T}_{\bar{\jmath}}} G_{\bar{T}_{\bar{\jmath}}} \ .
\end{equation}
Using eq.~\eqref{defY} we find that
\begin{equation}
 Y_{T_i  \bar{T}_{\bar{\jmath}}} = \frac{Y}{p}\left(-G_{T_i  \bar{T}_{\bar{\jmath}}} + \frac{1}{p} G_{T_i} G_{\bar{T}_{\bar{\jmath}}}\right) \ .
\end{equation}
Thus, we can compute
\begin{equation}
 Y_{T_i  \bar{T}_{\bar{\jmath}}} G^{T_i} = \frac{Y}{p^2} G_{\bar{T}_{\bar{\jmath}}} \left(G^{T_i} G_{T_i} - p \right)\ .
\end{equation}
From the above equation we read off that
\begin{equation}
 Y_{T_i  \bar{T}_{\bar{\jmath}}} G^{T_i} = 0 \ , \qquad \text{iff} \qquad G^{T_i} G_{T_i} = p \ ,
\end{equation}
which is nothing but the no-scale condition as defined in eq.~\eqref{eq:chiral_no_scale}. This completes the proof of the equivalence between solutions of the no-scale differential equation and the HCMA.
\section{Duality Transformation and Matching the Solutions}\label{duality_matching}
\subsection{Duality Transformation}\label{duality_transformation}
Let us discuss the duality transformation, which relates eq.~\eqref{Llinear} to a dual theory of chiral multiplets. We only reproduce the important formulae here, following \cite{Binetruy:2000zx, Grimm:2005fa}. Let us introduce the Lagrangian
\begin{equation}\label{L_Legendre_multiplier}
 \mathcal{L} = -3 \int E \left( F(L^i) + \tfrac{2}{3} L^i (T_i + \bar{T}_i) \right)+ \tfrac{1}{2} \int \frac{E}{R} \mathrm{e}^{K/2} W+ \tfrac{1}{2} \int \frac{E}{\bar{R}} \mathrm{e}^{K/2} \bar{W} \ ,
\end{equation}
where $L^i$ are only required to be real superfields.\footnote{Whenever we discuss models with a shift-symmetry we will drop the bars on the indices for the complex conjugate scalars.} Moreover, $K(L^i)$ denotes the K\"ahler potential and $W$ the constant superpotential. On one hand the equations of motion for $T_i$ derived from eq.~\eqref{L_Legendre_multiplier} are equivalent to eq.~\eqref{Def_linear_multi} and, hence, upon inserting the solution we retain the theory in eq.~\eqref{Llinear} \cite{Binetruy:2000zx}. On the other hand, varying eq.~\eqref{L_Legendre_multiplier} with respect to $L^i$ one finds \cite{Binetruy:2000zx, Grimm:2005fa}
\begin{equation}\label{T+Tb}
 \tfrac{2}{3}(T_i + \bar{T}_i) + F_{L^i} - \tfrac{1}{3} K_{L^i} (F+\tfrac{2}{3}L^j (T_j + \bar{T}_j)) = 0 \ .
\end{equation}
The above equation should be read as implicitly defining $L^i(T + \bar{T})$. 
The chiral theory is obtained after inserting $L^i(T + \bar{T})$ back into eq.~\eqref{L_Legendre_multiplier}. Since $L^i(T + \bar{T})$ is invariant under shifts $T_i - \bar{T}_i \rightarrow T_i - \bar{T}_i + C_i$ for some constants $C_i$, the dual chiral theory exhibits the looked-for Peccei-Quinn shift-symmetry. Moreover, the K\"ahler potential of the chiral theory is given by $K$ where again we replace $L^i$ with the solution of eq.~\eqref{T+Tb}. The Einstein-frame condition in eq.~\eqref{Einstein_frame} can now be recast into the form
\begin{equation}\label{F+TL=1}
 F(L^j) + \tfrac{2}{3}L^i (T_i + \bar{T}_i) = 1 \ .
\end{equation}
Up to a constant factor, which can be removed via a K\"ahler transformation, we can rewrite eq.~\eqref{kin_pot} as
\begin{equation}
 K(T+\bar{T}) = \tilde{K}(L(T+\bar{T})) - 2 (T_i + \bar{T}_i) L^i \ .
\end{equation}
This equation identifies $K$ to be the Legendre transform of $\tilde{K}$. Furthermore, the kinetic terms of $L^i$ in eq.~\eqref{L:linear} yield kinetic terms for the real parts of $T_i$ in the dual chiral theory. The kinetic terms for the imaginary component of $T_i$ are obtained by dualizing the three-form $H^i$ to $d \, \mathrm{Im}(T_i)$ by means of Hodge duality. Performing the explicit procedure yields the exact form of the component Lagrangian eq.~\eqref{L:chiral}, where the K\"ahler metric is derived from $K$, see \cite{Binetruy:2000zx, Grimm:2005fa}. 

Since we are interested in the scalar potential, we explicitly demonstrate the duality for $V$ given in eq.~\eqref{V:linear}. For brevity let us introduce the following notation for the real components of the chiral scalars
\begin{equation}\label{def:phi^alpha}
 \phi_i = \tfrac{1}{2} (T_i + \bar{T}_i) \ .
\end{equation}
Taking the derivative of eq.~\eqref{F+TL=1} with respect to $L^i$ and inserting eq.~\eqref{T+Tb} shows that
\begin{equation}\label{KLalpha=-4..}
 K_{L^i} = -4 L^j \frac{\partial \phi_j}{\partial L^i} \ .
\end{equation}
Multiplying this equation with the inverse matrix of derivatives we read off
\begin{equation}\label{Lalpha_Kprime}
 L^i = - \tfrac{1}{4} K_{\phi_i} \ .
\end{equation}
Taking derivatives with respect to $\phi_i$ we find
\begin{equation}\label{KTT}
 K_{\phi_i \phi_j} = -4 \frac{\partial  L^i}{\partial \phi_j} \ .
\end{equation}
Using eq.~\eqref{Lalpha_Kprime} and eq.~\eqref{KTT} we infer
\begin{equation}
 L^i K_{L^i} = L^i K_{\phi_j} \frac{\partial \phi_j}{\partial L^i} = K_{\phi_i}  K^{\phi_i \phi_j} K_{\phi_j} = K_{T_i}  K^{T_i \bar{T}_j} K_{\bar{T}_j}\ ,
\end{equation}
which shows that we obtain the correct scalar potential for a chiral theory with shift-symmetric K\"ahler potential.
\subsection{Matching the Solutions upon Dualization}\label{matching_the_sol}
%
%
Using the general prescription for the duality transformation, we now show that the solution given in eqs.~\eqref{sol_lin_1} and \eqref{sol_lin_2} matches the solution in eq.~\eqref{sol:chiral} after dualization to chiral multiplets.\footnote{It is worth pointing out that theories of chiral and linear multiplets are in general dual only at the classical level. The reason for this is that the linear multiplet does not include an auxiliary field and that off-shell higher-derivative corrections for chiral multiplets exist, which in particular correct the scalar potential by powers of the chiral auxiliary. For the special case of string-derived effective actions these corrections were recently discussed in \cite{Ciupke:2015msa}. Furthermore, the no-scale condition in these higher-derivative supergravities was studied in \cite{Cecotti:1986pk}.} 
First, replacing the K\"ahler potential via eq.~\eqref{k=-plogY} in eq.~\eqref{KLalpha=-4..} yields
\begin{equation}\label{Y_L}
 Y_{L^i} = \frac{4}{p} Y L^j \frac{\partial \phi_j}{\partial L^i} \ .
\end{equation}
Using this equation we find that
\begin{equation}
 \phi_i  Y_{\phi_i} = \phi_i Y_{L^j} \frac{\partial L^j}{\partial \phi_i}=\tfrac{4}{p} \phi_i  \, L^i Y \ .
\end{equation}
Inserting eq.~\eqref{F+TL=1} into the above we obtain
\begin{equation}
 \phi_i  Y_{\phi_i} = \tfrac{3}{p}(1-F)Y \ .
\end{equation}
Finally, using eqs.~\eqref{sol_lin_1} and \eqref{sol_lin_2} we conclude that
\begin{equation}\label{dualsol}
 Y = \phi_i  Y_{\phi_i} + \frac{3}{p}Y\left(\frac{L^2}{L^1},\dots,\frac{L^n}{L^1} \right) F_{(1)}\left(\frac{L^2}{L^1},\dots,\frac{L^n}{L^1} \right) \ ,
\end{equation}
where we have to read the equation as depending on the dual chiral coordinates. Comparing with eq.~\eqref{sol:chiral} and recalling the definition of the real variables in eq.~\eqref{def:phi^I} we find that the solutions match, after identifying
\begin{equation}\label{k_linear}
  \tilde{Y}\left(\frac{L^2}{L^1},\dots,\frac{L^n}{L^1} \right) = \frac{3}{p}Y\left(\frac{L^2}{L^1},\dots,\frac{L^n}{L^1} \right) F_{(1)}\left(\frac{L^2}{L^1},\dots,\frac{L^n}{L^1} \right) \ ,
\end{equation}
and the $n-1$ coordinates $u_\alpha$ are related to the real scalars inside the linear multiplets via
\begin{equation}\label{u^A}
 u_\alpha = \frac{L^\alpha}{L^1} \ , \qquad \alpha = 1,\dots,n-1 \ ,
\end{equation}
which again has to be read as an equality of functions each depending on $(T_i + \bar{T}_i)$. It remains to be seen that eq.~\eqref{sol1} holds, i.e.~we have to see whether $Y_{T_i}$ can be expressed as a function of the variables $u_\alpha$. Replacing $K$ via eq.~\eqref{k=-plogY} in eq.~\eqref{Lalpha_Kprime} we observe
\begin{equation}\label{Y_T}
 Y_{\phi_i} = \tfrac{4}{p} Y L^i \ .
\end{equation}
Since $Y$ is homogeneous of degree -1 in the $L^i$, we can express the above as
\begin{equation}\label{gi_linear}
 Y_{\phi_i} = \frac{4}{p} \frac{L^i}{L^1} Y\left(\frac{L^2}{L^1},\dots,\frac{L^n}{L^1}\right) \equiv g^i \left(\frac{L^2}{L^1},\dots,\frac{L^n}{L^1} \right) \ ,
\end{equation}
which is indeed a function of the variables $u_\alpha$ according to eq.~\eqref{u^A}.


Let us finish this section by checking that the constraint equations in eq.~\eqref{sol3} are automatically satisfied upon dualization. 
More precisely this amounts to the vanishing of the following object
\begin{equation}\label{C_A}
 C^{\alpha}(\phi) = \phi_i \,\partial_{u_\alpha} g^i(u) + \partial_{u_\alpha} \tilde{Y}(u) \ ,
\end{equation}
where the $u_\alpha$ are understood as functions of the dual chiral fields $\phi_i$. To check that $C^\alpha$ vanish we must replace $\tilde{Y}(u_\alpha)$ and $g^i(u_\alpha)$ via eqs.~\eqref{k_linear} and \eqref{gi_linear} in eq.~\eqref{C_A}. 
Before performing this replacement we rewrite $\tilde{Y}$ and $g^i$ as functions of $L^i$.\footnote{From now on we omit writing the dependence upon the $\phi_i$ but it is implied.} Using eq.~\eqref{sol_lin_1} for $Y(L)$ and similarly that $F_{(1)}(L)$ is a homogeneous function of degree one, we obtain 
\begin{equation}
\tilde{Y}(u_\alpha)=\tfrac{3}{p} Y(L) F_{(1)}(L) \quad \mathrm{and} \quad  g^i(u_\alpha) = \tfrac{4}{p} L^i Y(L) \ .
\end{equation} 
After replacing these in eq.~\eqref{C_A} and using the chain rule for derivatives we find
\begin{equation}
C^\alpha (\phi) = \left(\, \tfrac{4}{p} \phi_i Y(L) + \tfrac{4}{p} \phi_j L^j Y_{L^i}(L) + \tfrac{3}{p}Y_{L^i}(L)F_{(1)}(L) + \tfrac{3}{p}Y(L)F_{(1)L^i}(L)\,\right)\frac{\partial L^i}
{\partial u_\alpha} \ .
\end{equation}
Furthermore, using eqs.~\eqref{sol_lin_2} and \eqref{F+TL=1} the above reduces to
\begin{equation}
C^\alpha (\phi) = \left(\,\tfrac{4}{p}\phi_i Y(L)+Y_{L^i}(L) + \tfrac{3}{p}Y(L)F_{(1)L^i}(L)\,\right)\frac{\partial L^i}
{\partial u_\alpha} \ .
\end{equation}
Finally, by means of eqs.~\eqref{T+Tb}, \eqref{F+TL=1} and \eqref{k=-plogY} we directly obtain that the terms inside the parenthesis cancel. Thus, as promised $C^\alpha$ vanish.
\bibliographystyle{JHEP}
\bibliography{CZ_final_v2}

\end{document}